\documentclass[11pt]{article}
\usepackage[letterpaper,margin=1.02in]{geometry}
\usepackage{microtype,placeins}
\usepackage{tikz,pgfplots}
\usetikzlibrary{arrows.meta,decorations.pathreplacing}
\usepgfplotslibrary{groupplots}
\pgfplotsset{compat=1.18,every axis/.append style={
  tick label style={font=\scriptsize},label style={font=\small},
  title style={font=\small\bfseries},legend style={font=\scriptsize,draw=none},
  tick align=outside,grid=major,major grid style={black!10},
  every axis plot/.append style={line width=.8pt}}}
\newenvironment{acknowledgments}{\section*{Acknowledgments}}{}
\usepackage[utf8]{inputenc}
\usepackage{amsmath,amssymb,amsthm,braket,graphicx}
\usepackage[colorlinks=true,linkcolor=blue,citecolor=blue,urlcolor=blue]{hyperref}
\newtheorem{theorem}{Theorem}
\newtheorem{proposition}{Proposition}
\newtheorem{lemma}{Lemma}
\newtheorem{corollary}{Corollary}
\theoremstyle{definition}
\newcommand{\one}{\mathbf{1}}
\newcommand{\tr}{\operatorname{Tr}}
\newcommand{\E}{\mathbb{E}}
\newcommand{\bb}{\mathbf{b}}
\newcommand{\DEN}{D_{\mathrm{EN}}}

\begin{document}
\raggedbottom

\title{AKLT Entanglement via Quantum Filtering}
\author{Amenallah Andolsi\\
\small Nanostructured Materials, Quantum and Nonlinear Optics Laboratory\\
\small Faculty of Sciences of Tunis, Tunis El Manar University, Tunis, Tunisia\\
\small \href{mailto:amenallah.andolsi@fst.utm.tn}{amenallah.andolsi@fst.utm.tn}}
\date{}
\maketitle

\begin{abstract}
Local measurements on the spin-1 Affleck--Kennedy--Lieb--Tasaki (AKLT) chain can preserve entanglement between distant parties when the outcomes are retained. We describe this protocol through the hidden quantum Markov filter on the virtual qubit and compare two choices of target systems: the two edge qubits of an open segment and two physical spins in the infinite chain. For the edges, a determinant identity gives an exact basis-dependent decay law for average concurrence and negativity. The conditional entanglement entropy and the Holevo information carried by the record sum to one bit. For physical spins, the record-free state is separable beyond nearest neighbors, although its mutual information has a nonzero leading term of order $9^{-r}$. Cartesian measurements make every conditional spin pair locally unitarily equivalent to the nearest-neighbor state. Measurements in the $S^z$ basis instead suppress average negativity by a factor of $3^{-(r-1)}$. At distance two, a non-Cartesian projective basis gives an exact average negativity greater than the nearest-neighbor value. We also relate Ohya's degree of entanglement to the hashing bound and show how its sign distinguishes the edge ensembles from the entangled physical-spin examples considered here.
\end{abstract}

\section{Introduction}\label{sec:introduction}
In a spin chain, the distance over which two sites remain correlated can differ sharply from the distance over which measurements can distribute entanglement. The spin-1 Affleck--Kennedy--Lieb--Tasaki (AKLT) chain gives an exact example. Affleck, Kennedy, Lieb, and Tasaki constructed the valence-bond ground state~\cite{Affleck1987,Affleck1988}. It admits a matrix product representation of bond dimension two~\cite{Fannes1992,Cirac2021}. The three nontrivial eigenvalues of its transfer channel are all $-1/3$, giving a bulk correlation length of $1/\ln 3$ in lattice units. Yet local measurements on an arbitrarily long segment can prepare a Bell pair between its edge qubits, provided the outcomes are retained and used to correct the edge state~\cite{VerstraeteMartinDelgadoCirac2004,Popp2005}. This distinction underlies localizable entanglement, which studies the average entanglement obtainable between chosen targets by measuring other parts of a system~\cite{VerstraetePoppCirac2004,Popp2005}.

The AKLT chain also illustrates how this measurement procedure becomes a primitive for quantum information processing. Its virtual spin-$\tfrac12$ degrees of freedom carry a projective action of the protecting spin rotations~\cite{Pollmann2010,Pollmann2012}. In a Cartesian spin basis, a local readout applies a Pauli operator to the virtual qubit, with the identity of that operator determined by the outcome. The outcome string therefore supplies the information needed to track or correct the net Pauli byproduct. This is the quantum-wire mechanism developed in matrix-product approaches to measurement-based quantum computation~\cite{GrossEisert2007,BrennenMiyake2008}. Subsequent work placed such wire protocols in the broader setting of symmetry-protected phases~\cite{Else2012,Raussendorf2017}. The protocol analyzed here uses a fixed projective basis and classical communication of the outcomes (LOCC)~\cite{Chitambar2014,Horodecki2009}. This restriction lets us compare bases through their full outcome ensembles.

A useful description of the protocol comes from quantum filtering. Belavkin's filtering theory gives the conditional state of a quantum system given a classical observation record~\cite{Belavkin1992,Bouten2007}. Monras, Beige, and Wiesner connected hidden quantum Markov models (HQMMs) to the nonadaptive readout of many-body states~\cite{Monras2011}. In the AKLT setting, Souissi and Andolsi identified the spin-1 chain as the output of a hidden spin-$\tfrac12$ system and related the symmetry-protected order to covariance of the decoding channels~\cite{SouissiAndolsiEPJP2026}. Their subsequent causal HQMM preprint gives an observation-process representation of the infinite AKLT state at the level of finite-block density matrices~\cite{SouissiAndolsi2026causal}. In the present protocol, a projective readout converts those quantum outputs into classical outcomes. The resulting instrument gives both the Born probability of a record and the conditional virtual state. Summing over outcomes recovers the same transfer channel for every orthonormal measurement basis, while keeping the record exposes how the individual Kraus products depend on the chosen basis.

The distinction between virtual edges and physical targets is important in interpreting earlier entanglement calculations. Fan, Korepin, and Roychowdhury derived the entropy of a contiguous block in the valence-bond state~\cite{FanKorepinRoychowdhury2004}. Santos, Korepin, and Bose studied negativity for separated AKLT blocks, and Santos and Korepin also analyzed mutual entropy in this geometry~\cite{SantosKorepinBose2011,SantosKorepin2012}. Those results concern reduced states obtained by tracing out other spins. For the physical-pair geometry studied here, we use the entanglement-breaking property of the virtual transfer channel to establish separability as soon as an unobserved site lies between the targets. Conditioning on measurements of the intervening sites yields a different ensemble, whose members are generally mixed. The nearest-neighbor state then provides a common reference for comparing assisted protocols at different separations.

We obtain three quantitative results. First, for the two edge qubits, a determinant identity gives the exact average concurrence and negativity in an arbitrary orthonormal basis. It also bounds the mean entanglement entropy. The Holevo information of the classical record and Bob's edge adds to that mean entropy to give one bit. Second, for physical spin pairs, the discarded-record mutual information decays with an explicit leading coefficient even though the state is separable beyond nearest neighbors. Keeping a Cartesian measurement record makes every conditional pair locally unitarily equivalent to the nearest-neighbor state. Third, the basis preferred for the edge protocol and the basis preferred for physical-pair negativity can differ: at separation two we give an exact non-Cartesian example whose mean negativity exceeds the Cartesian value. Negativity and concurrence quantify different aspects of the conditional states~\cite{VidalWerner2002,Wootters1998}. We also compare negativity with the entropic degree of entanglement used in studies of entangled Markov chains~\cite{AccardiMatsuokaOhya2006,AccardiSouissiDEN2025} and relate that degree to the directional hashing bounds~\cite{DevetakWinter2005}.

These comparisons also suggest concrete benchmarks for hardware. Smith \emph{et al.} demonstrated AKLT preparation on a quantum processor using fusion measurements~\cite{Smith2023}. Kumaran \emph{et al.} subsequently developed a transmon-qutrit implementation for simulating spin-1 AKLT systems~\cite{Kumaran2026}. The protocols in Fig.~\ref{fig:setup} isolate a different experimental task: measuring selected spin-1 outputs, retaining the classical record, and comparing the resulting endpoint states across bases. The edge protocol predicts a basis-dependent concurrence and a known Pauli correction in the Cartesian setting. The physical-spin protocol predicts a measurable change in Born-weighted negativity at separation two. Both predictions follow from the ideal AKLT tensors and can serve as reference values for record-resolved state reconstruction.

Section~\ref{sec:aklt-filter} introduces the transfer channel and measurement instrument. Section~\ref{sec:measures} specifies the two protocols and their observables. Sections~\ref{sec:discarded-record} and~\ref{sec:retained-record} give the results with the record discarded and retained, respectively. Section~\ref{sec:discussion} discusses their implications for entanglement distribution.

\section{AKLT tensors, transfer channel, and quantum filtering}\label{sec:aklt-filter}
\subsection{Matrix product representation and edge symmetry}
The spin-1 AKLT Hamiltonian is
\begin{equation}
  H_{\mathrm{AKLT}} = \sum_i \Big[\vec S_i\cdot\vec S_{i+1}+\tfrac{1}{3}(\vec S_i\cdot\vec S_{i+1})^2\Big].
\end{equation}
The translation-invariant infinite-chain valence-bond ground state is a finitely correlated state (MPS) of bond dimension $D=2$~\cite{Fannes1992,PerezGarcia2007,Cirac2021}, with tensors on the virtual space $\mathcal H_V\simeq\mathbb C^2$
\begin{equation}
  A_+=\sqrt{\tfrac23}\,\sigma^+,\quad A_0=-\tfrac{1}{\sqrt3}\,\sigma^z,\quad A_-=-\sqrt{\tfrac23}\,\sigma^-,
  \label{eq:A}
\end{equation}
where $\sigma^\pm=(\sigma^x\pm i\sigma^y)/2$. They satisfy $\sum_kA_k^\dagger A_k=\sum_kA_kA_k^\dagger=\one$. The transfer map $\mathcal T(X)=\sum_kA_kXA_k^\dagger$ obeys
\begin{equation}
  \mathcal T(\one)=\one,\qquad \mathcal T(\sigma^a)=\epsilon\,\sigma^a,\qquad \epsilon=-\tfrac13,
\end{equation}
so all three subleading eigenvalues equal $-\tfrac13$ and $\xi=-1/\ln|\epsilon|=1/\ln3$~\cite{Fannes1992,Schollwoeck2011}. The tensors form a spin-1 vector operator,
\begin{equation}
  u_gA_ku_g^\dagger=\sum_{k'}D_{k'k}(g)A_{k'},\qquad g\in SO(3),
  \label{eq:cov}
\end{equation}
where $u_g\in SU(2)$ is a lift of $g$ and $D(g)$ is the spin-1 representation in the tensor convention of Eq.~\eqref{eq:cov}. The edge therefore carries a \emph{projective} representation of the protecting symmetry. For $g\in D_2=\{\one,R_x(\pi),R_y(\pi),R_z(\pi)\}$, its nonidentity elements lift to Pauli matrices up to phase, and distinct Pauli matrices anticommute; this is the hallmark of the Haldane SPT phase~\cite{Pollmann2010,Pollmann2012}. The projective edge representation is stable under symmetry-preserving gapped deformations within the phase~\cite{Pollmann2010,Pollmann2012}. The exact transfer eigenvalue and the measurement formulas below refer to the AKLT point.

\subsection{Unconditional dynamics of the virtual qubit}
When the physical output is traced out, the transfer map is the channel $\mathcal E=\mathcal T$ on the hidden qubit~\cite{SouissiAndolsiEPJP2026,Monras2011}. It fixes the maximally mixed state and multiplies every Bloch-vector component by $\epsilon$. After $n$ sites its action is
\begin{equation}
  \mathcal E^n(X)=\eta_nX+(1-\eta_n)\tr(X)\tfrac{\one}{2}.
  \label{eq:depol}
\end{equation}

\subsection{The AKLT state as an HQMM observation process}
Souissi and Andolsi give a causal HQMM realization of the AKLT state in which the physical spin chain is the quantum observation process~\cite{SouissiAndolsi2026causal}. Their published AKLT framework describes the emission from a virtual spin-$\tfrac12$ system and its relation to the symmetry of the chain~\cite{SouissiAndolsiEPJP2026}. We use the matrix product tensors of Eq.~\eqref{eq:A} to write the observation process directly in the Schr\"odinger picture and then derive the classical probability law for a specified readout.

The isometry $W=\sum_k\ket{k}\otimes A_k$ emits a physical spin-1 and passes the virtual qubit to the next site. For a block of $m$ emitted spins, put $W_1=W$ and $W_{m+1}=(\one_{3^m}\otimes W)W_m$, with the virtual output ordered last. The quantum observation on that block is the state
\[
  \Gamma_m=\operatorname{Tr}_{V}\!\left[W_m\,(\one_2/2)\,W_m^\dagger\right].
\]
Indeed, its matrix element at strings $\mathbf k=(k_1,\dots,k_m)$ and $\mathbf l=(l_1,\dots,l_m)$ is
\[
  \langle\mathbf k|\Gamma_m|\mathbf l\rangle
  =\tfrac12\tr\!\left[A_{k_m}\cdots A_{k_1}
      A_{l_1}^{\dagger}\cdots A_{l_m}^{\dagger}\right],
\]
the canonical AKLT block formula with left virtual density $\one_2/2$ and right boundary $\one_2$. The isometry relation $W^\dagger W=\one_2$ gives consistency when the last emitted spin is traced out, while $\mathcal E(\one_2/2)=\one_2/2$ gives consistency under a shift of the block by one site. The family $(\Gamma_m)$ thus describes the stationary quantum output before a measurement basis is specified.

\subsection{Measurement instrument and conditional virtual dynamics}
Measuring each emitted spin in an orthonormal basis $\bb=\{\ket{b_j}\}_{j=1}^3$ defines the instrument shown in Fig.~\ref{fig:setup}(a):
\begin{equation}
\begin{aligned}
  &K^\bb_j=\sum_k\braket{b_j|k}A_k,\\
  &\sum_jK^{\bb\dagger}_jK^\bb_j=\one,\qquad \sum_jK^\bb_j(\cdot)K^{\bb\dagger}_j=\mathcal E,
\end{aligned}
  \label{eq:Kb}
\end{equation}
so every basis unravels the same channel. Given a positive-probability record $\mathcal D_n=(j_1,\dots,j_n)$ and the product $K_{\mathcal D}=K_{j_n}\cdots K_{j_1}$, the HQMM (quantum Bayes) filter is
\begin{equation}
  \rho_n(\mathcal D_n)=\frac{K_{\mathcal D}\rho_0K_{\mathcal D}^\dagger}{p(\mathcal D_n)},\qquad p(\mathcal D_n)=\tr\bigl(K_{\mathcal D}\rho_0K_{\mathcal D}^\dagger\bigr).
  \label{eq:filter}
\end{equation}
The measurement makes the observation process classical. For the stationary initialization $\rho_0=\one_2/2$, its finite-record probabilities are the diagonal matrix elements of $\Gamma_n$ in the chosen product basis:
\[
  \Pr(J_1=j_1,\dots,J_n=j_n)
  =\langle b_{j_1}\otimes\cdots\otimes b_{j_n}|
       \Gamma_n|b_{j_1}\otimes\cdots\otimes b_{j_n}\rangle
  =p(\mathcal D_n).
\]
The last equality follows by projecting each emitted spin in $W_n$, which leaves the virtual product $K_{\mathcal D}$. Summing over the last outcome recovers the probability of the shorter record because $\sum_jK_j^\dagger K_j=\one_2$. In particular, the Born probabilities define a consistent law on outcome sequences. For any initial $\rho_0$ and any positive-probability history $\mathcal D_n$, the ratio of successive record probabilities is
\[
  \Pr(J_{n+1}=j\mid\mathcal D_n)=\tr(K_j\rho_n(\mathcal D_n)K_j^\dagger).
\]
When this probability is positive, the normalized operator $K_j\rho_n(\mathcal D_n)K_j^\dagger$ divided by its trace is the updated virtual density. The record can have temporal correlations, but its effect on the next prediction is summarized by the conditional density $\rho_n(\mathcal D_n)$. This is the probabilistic role of the hidden qubit in the entanglement calculations below. In the edge setup it gives a reduced density matrix of the conditional pure edge pair, and in the physical-spin setup the same Kraus product enters the conditional two-site state.

Equation~\eqref{eq:filter} is the discrete-time Belavkin filter~\cite{Belavkin1992,Bouten2007} and the forward recursion of the HQMM~\cite{Monras2011,Srinivasan2018}. Its trajectories purify whenever no dark subspace exists~\cite{MaassenKummerer2006,Benoist2019}, at a rate recently bounded in general in Ref.~\cite{Bompais2026}. For the real family $\ket{b_1}=-\cos\theta\ket{+}+\sin\theta\ket{-}$, $\ket{b_0}=\ket0$, $\ket{b_2}=\sin\theta\ket{+}+\cos\theta\ket{-}$, the value $\theta=0$ gives the $S^z$ basis and $\theta=\pi/4$ the Cartesian basis $\{\ket x,\ket y,\ket z\}$, for which $K_a\propto\sigma^a$~\cite{GrossEisert2007,BrennenMiyake2008}.

The following determinant identity makes the average edge entanglement accessible in closed form. It is included with proof to make the analysis self-contained.

\begin{lemma}[Exact determinant law]\label{lem:law}
Let $\mathcal M(\rho)=2\sqrt{\det\rho}$. For every basis $\bb$ and every $\rho_0$,
\begin{equation}
  \E\bigl[\mathcal M(\rho_n)\bigr]=\mathcal M(\rho_0)\,s_\bb^{\,n},\quad s_\bb=\tfrac13\sum_j|\bb_j\cdot\bb_j|=e^{-\kappa_\bb},
  \label{eq:law}
\end{equation}
where $b\cdot b=b_0^2-2b_+b_-$ is the invariant bilinear form in the spherical basis. One has $1/3\le s_\bb\le1$. For the real family on $0\le\theta\le\pi/2$, $s(\theta)=(1+2\sin 2\theta)/3$.
\end{lemma}

\begin{proof}
Since $\det$ is multiplicative, $\mathcal M(\rho_n)\,p(\mathcal D_n)=\mathcal M(\rho_0)\prod_i|\det K_{j_i}|$. A direct computation from Eq.~\eqref{eq:A} gives $|\det K^\bb_j|=|\bb_j\cdot\bb_j|/3$. A record of zero probability has a zero contribution to the determinant identity, so summing gives Eq.~\eqref{eq:law}. In Cartesian coordinates the matrix $Q$ with entries $Q_{jk}=b_j\cdot b_k$ is symmetric unitary. The upper bound follows from $|Q_{jj}|\le1$. For the lower bound set $Z=Q-\operatorname{diag}(Q_{11},Q_{22},Q_{33})$. A diagonal unitary congruence makes its three off-diagonal entries real and nonnegative, preserving its singular values. The resulting real matrix has trace zero and determinant $2Z'_{12}Z'_{13}Z'_{23}\ge0$, so its singular values satisfy $z_1=z_2+z_3$ in decreasing order, including the singular case. Mirsky's trace-norm singular-value perturbation inequality~\cite{Mirsky1960} gives
\[
  \sum_j|Q_{jj}|=\|Q-Z\|_1\ge\sum_j|1-z_j|
    \ge |(1-z_2)+(1-z_3)-(1-z_1)|=1.
\]
Substitution into the spherical-coordinate formula yields the stated dependence on $\theta$.
\end{proof}

\section{LOCC setting and entanglement measures}
\label{sec:measures}
\subsection{Two LOCC scenarios}
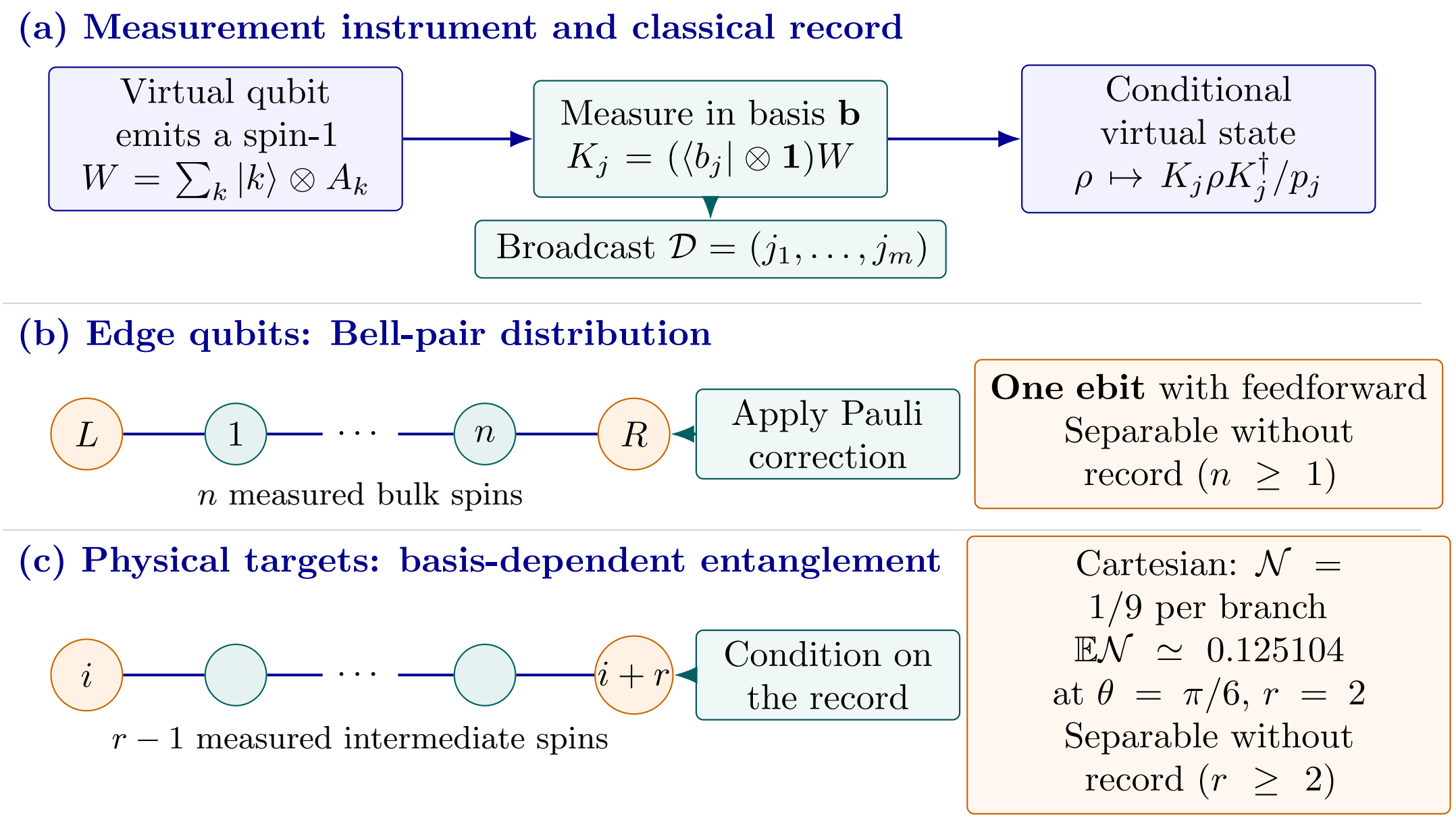
\begin{figure}[htbp]
\centering
\begin{tikzpicture}[x=1cm,y=1cm,>=Latex,
  every node/.style={font=\small},
  heading/.style={anchor=west,font=\small\bfseries,text=blue!55!black},
  target/.style={circle,draw=orange!80!black,fill=orange!11,minimum size=7mm,inner sep=0pt},
  site/.style={circle,draw=teal!75!black,fill=teal!10,minimum size=6mm,inner sep=0pt},
  block/.style={draw=teal!65!black,fill=teal!6,rounded corners=2pt,
    align=center,text width=3.25cm,minimum height=1.14cm,inner sep=3pt},
  input/.style={draw=blue!55!black,fill=blue!5,rounded corners=2pt,
    align=center,text width=3.25cm,minimum height=1.14cm,inner sep=3pt},
  result/.style={draw=orange!78!black,fill=orange!6,rounded corners=2pt,
    align=center,text width=4.3cm,minimum height=1.08cm,inner sep=4pt},
  qwire/.style={draw=blue!57!black,line width=.75pt},
  classical/.style={draw=teal!75!black,densely dashed,line width=.8pt,->}]

\node[heading] at (.12,.1) {(a) Measurement instrument and classical record};
\node[input] (emission) at (2.30,-.97) {Virtual qubit emits a spin-1\\
  $W=\sum_k\ket{k}\otimes A_k$};
\node[block] (measure) at (7.05,-.97) {Measure in basis $\bb$\\
  $K_j=(\bra{b_j}\otimes\one)W$};
\node[input] (update) at (11.83,-.97) {Conditional virtual state\\
  $\rho\mapsto K_j\rho K_j^\dagger/p_j$};
\draw[qwire,->] (emission) -- (measure);
\draw[qwire,->] (measure) -- (update);
\node[block,text width=4.4cm,minimum height=.48cm] (flag) at (7.05,-2.05)
  {Broadcast $\mathcal D=(j_1,\ldots,j_m)$};
\draw[classical] (measure.south) -- (flag.north);

\draw[gray!34] (.12,-2.58) -- (14.01,-2.58);
\node[heading] at (.12,-2.90) {(b) Edge qubits: Bell-pair distribution};
\node[target] (left) at (.94,-3.86) {$L$};
\node[site] (s1) at (2.4,-3.86) {$1$};
\node (sd) at (3.62,-3.86) {$\cdots$};
\node[site] (sn) at (4.84,-3.86) {$n$};
\node[target] (right) at (6.30,-3.86) {$R$};
\draw[qwire] (left)--(s1) (s1)--(sd) (sd)--(sn) (sn)--(right);
\node[font=\footnotesize] at (3.62,-4.47) {$n$ measured bulk spins};
\node[block,text width=2.37cm,minimum height=.77cm] (feed) at (8.20,-3.86)
  {Apply Pauli\\correction};
\draw[classical] (feed.west) -- (right.east);
\node[result] at (11.93,-3.86) {\textbf{One ebit} with feedforward\\
  Separable without record ($n\ge1$)};

\draw[gray!34] (.12,-4.80) -- (14.01,-4.80);
\node[heading] at (.12,-5.12) {(c) Physical targets: basis-dependent entanglement};
\node[target] (p) at (.94,-6.22) {$i$};
\node[site] (u) at (2.4,-6.22) {};
\node (ud) at (3.62,-6.22) {$\cdots$};
\node[site] (v) at (4.84,-6.22) {};
\node[target] (q) at (6.30,-6.22) {$i+r$};
\draw[qwire] (p)--(u) (u)--(ud) (ud)--(v) (v)--(q);
\node[font=\footnotesize] at (3.62,-6.86) {$r-1$ measured intermediate spins};
\node[block,text width=2.37cm,minimum height=.77cm] (known) at (8.20,-6.22)
  {Condition on\\the record};
\draw[classical] (known.west) -- (q.east);
\node[result,text width=4.45cm,minimum height=1.48cm] at (11.93,-6.22)
  {Cartesian: $\mathcal N=1/9$ per branch\\
   $\E\mathcal N\simeq0.125104$\\
   at $\theta=\pi/6$, $r=2$\\
   Separable without record ($r\ge2$)};
\end{tikzpicture}

\caption{Measurement, classical feedforward, and entanglement distribution in the AKLT chain. (a) The emission isometry $W$ and the measurement basis determine the conditional update of the virtual qubit. Here $p_j=\tr(K_j\rho K_j^\dagger)$ and the displayed update applies for $p_j>0$. Dashed arrows indicate classical communication. (b) For Cartesian measurements, the record specifies a Pauli byproduct on the right edge. Applying its inverse gives a common Bell state, up to the fixed choice of edge basis. The discarded-record state is separable for $n\ge1$. (c) Between physical targets, the record labels a generally mixed conditional state. The Cartesian protocol gives the nearest-neighbor negativity at every distance. At distance two, the real-family angle $\theta=\pi/6$ gives a larger Born-weighted mean.}
\label{fig:setup}
\end{figure}

\emph{Edge--edge.} Consider an open AKLT segment of $n$ sites whose left dangling spin-$\tfrac12$ $L$ (Alice) forms a singlet with the first virtual spin. Bob holds the right dangling spin-$\tfrac12$ $R$, and Charlie holds the $n$ bulk spin-1's [Fig.~\ref{fig:setup}(b)]. Up to a local unitary, the state of $LR$ given Charlie's record is $(\one\otimes K_{\mathcal D})\ket\Phi/\sqrt{p}$, where $\ket\Phi$ is a normalized Bell vector and $\Phi=\ket\Phi\!\bra\Phi$. Its reduced state on $R$ is the filter state $\rho_n(\mathcal D_n)$ with $\rho_0=\one/2$. If Charlie's record is discarded, the state is the Choi state
\begin{equation}
  J_n=(\mathrm{id}\otimes\mathcal E^n)(\Phi)=\eta_n\Phi+(1-\eta_n)\tfrac{\one}{4}.
  \label{eq:choi}
\end{equation}
Measuring and broadcasting is LOCC for the tripartition $L|C|R$. In particular, for the cut $L|CR$ it is LOCC on Alice's partner. The initial state is pure across $L|CR$ and has $S(L)=1$. The average conditional entanglement entropy of $LR$ is bounded by this one-ebit cut entropy.

\emph{Spin--spin.} Take two physical spins $i$ and $i+r$ in the infinite chain. The $r-1$ spins between them are measured in $\bb$ and broadcast, and all other spins are traced out [Fig.~\ref{fig:setup}(c)]. With left and right boundary operators $\one/2$ and $\one$, the unnormalized conditional two-spin state is
\begin{equation}
  \widetilde\rho_{i,i+r}^{(\mathcal D)}(s,t;s',t')=\tr\bigl[A_t\,K_{\mathcal D}A_s\tfrac{\one}{2}A_{s'}^\dagger K_{\mathcal D}^\dagger\,A_{t'}^\dagger\bigr].
  \label{eq:pair}
\end{equation}
Summing the diagonal indices and using both canonical identities gives $p(\mathcal D)=\tr(K_{\mathcal D}K_{\mathcal D}^\dagger)/2$. Thus $\rho^{(\mathcal D)}=\widetilde\rho^{(\mathcal D)}/p(\mathcal D)$ on positive-probability branches. Replacing $K_{\mathcal D}(\cdot)K_{\mathcal D}^\dagger$ by $\mathcal E^{r-1}$ sums the unnormalized states and yields the record-free pair. For $r=1$ the sole record is empty and its product is $\one$.

\subsection{Entanglement and correlation measures}
For a bipartite state $\theta$ on $A\otimes B$ with marginals $\rho_A,\rho_B$, we use the following quantities (logarithms base 2).

\emph{Quantum mutual information}, $I(A{:}B)=S(\rho_A)+S(\rho_B)-S(\theta)=S(\theta\Vert\rho_A\otimes\rho_B)$, which quantifies total (classical plus quantum) correlations~\cite{GroismanPopescuWinter2005}. In Ohya's language it is the quasi-mutual entropy $I_\theta(\rho_A,\rho_B)$ of the compound state $\theta$~\cite{BelavkinOhya2002}.

\emph{Negativity and logarithmic negativity}~\cite{Peres1996,VidalWerner2002,Plenio2005}, $\mathcal N=(\Vert\theta^{T_B}\Vert_1-1)/2$ and $E_{\mathcal N}=\log_2\Vert\theta^{T_B}\Vert_1$. These are LOCC monotones that vanish on PPT states, and PPT states are undistillable~\cite{Horodecki1998}.

\emph{Concurrence}~\cite{Wootters1998}, defined for two qubits. For a pure two-qubit state, let $\lambda_1,\lambda_2$ be the eigenvalues of $\rho_A$ (the squared Schmidt coefficients). Then $C=2\sqrt{\lambda_1\lambda_2}=\mathcal M(\rho_A)$ and $\mathcal N=C/2$.

\emph{Ohya's degree of entanglement}~\cite{BelavkinOhya2002,AccardiMatsuokaOhya2006},
\begin{equation}
  \DEN(\theta;\rho_A,\rho_B)=\tfrac12\bigl[S(\rho_A)+S(\rho_B)\bigr]-I_\theta(\rho_A,\rho_B).
  \label{eq:den}
\end{equation}
For pure states, $\DEN=-S(\rho_A)$, which is negative if and only if the state is entangled~\cite{AccardiMatsuokaOhya2006}. The following lemma gives $\DEN$ an operational meaning for mixed states.

\begin{lemma}\label{lem:den}
(i) $\DEN=-\tfrac12\bigl[I_c(A\rangle B)+I_c(B\rangle A)\bigr]$, where $I_c(A\rangle B)=S(\rho_B)-S(\theta)$ is the coherent information. (ii) If $\theta$ is separable, then $\DEN\ge0$. (iii) For fixed $A\to B$ communication, $E_D^{A\to B}(\theta)\ge\max\{0,I_c(A\rangle B)\}$. Allowing a choice of direction gives $\max\{E_D^{A\to B},E_D^{B\to A}\}\ge\max\{0,-\DEN(\theta)\}$.
\end{lemma}

\begin{proof}
(i) follows from Eq.~\eqref{eq:den} and $I=S_A+S_B-S_{AB}$. (ii) Separable states satisfy $S(\theta)\ge\max\{S(\rho_A),S(\rho_B)\}$~\cite{NielsenKempe2001}, so both coherent informations are nonpositive. (iii) applies the hashing inequality separately in each direction~\cite{DevetakWinter2005} and uses $\max\{x,y\}\ge(x+y)/2$.
\end{proof}

Hence $\DEN<0$ certifies \emph{distillable} entanglement via hashing, whereas $\DEN\ge0$ is inconclusive for mixed states. The quantity $\DEN$ is the Ohya counterpart of the hashing bound, which explains why it is natural in quantum-Markov-chain settings~\cite{AccardiMatsuokaOhya2006,AccardiSouissiDEN2025}.

\section{Entanglement structure without the record}\label{sec:discarded-record}
\subsection{Blocks and edges}
The reduced state of a block of $\ell$ consecutive spins has the same nonzero spectrum as $J_\ell$. Indeed, the block-emission isometry applied to one half of a Bell pair purifies both reduced states on complementary subsystems. This argument includes $\ell=1$, where $J_\ell$ has rank three. Equation~\eqref{eq:choi} therefore gives its spectrum $\{\tfrac{1+3\eta_\ell}{4},\tfrac{1-\eta_\ell}{4}{\times}3\}$, reproducing the block entropy $S_\ell\to2$ bits of Ref.~\cite{FanKorepinRoychowdhury2004}.

\begin{proposition}[Record-free edges]\label{prop:edges}
For $n\ge1$ the edge state $J_n$ is separable, so $\mathcal N=E_{\mathcal N}=0$. Moreover,
\begin{equation}
  I(L{:}R)=2-S(J_n),\qquad \DEN(J_n)=S(J_n)-1\ \ge0,
\end{equation}
with $S(J_n)=H\bigl(\tfrac{1+3\eta_n}{4},\tfrac{1-\eta_n}{4},\tfrac{1-\eta_n}{4},\tfrac{1-\eta_n}{4}\bigr)$.
\end{proposition}

\begin{proof}
The partial transpose of $J_n$ has eigenvalues $\tfrac{1+\eta_n}{4}$ (three times) and $\tfrac{1-3\eta_n}{4}$, all nonnegative for $-\tfrac13\le\eta_n\le\tfrac13$. For two qubits, PPT implies separability~\cite{Horodecki1996}. Equivalently, the channel $\mathcal E$ is entanglement breaking because $|\epsilon|=\tfrac13$~\cite{Horodecki2003}. The marginals are maximally mixed, which gives the formulas for $I$ and $\DEN$.
\end{proof}

At $n=1$, $J_1=(\one-\Phi)/3$ has mutual information $I=2-\log_23\approx0.415$ bits despite being separable. As $n\to\infty$, $I(L{:}R)\to0$ and $\DEN\to1$.

\subsection{Pairs of physical spins}
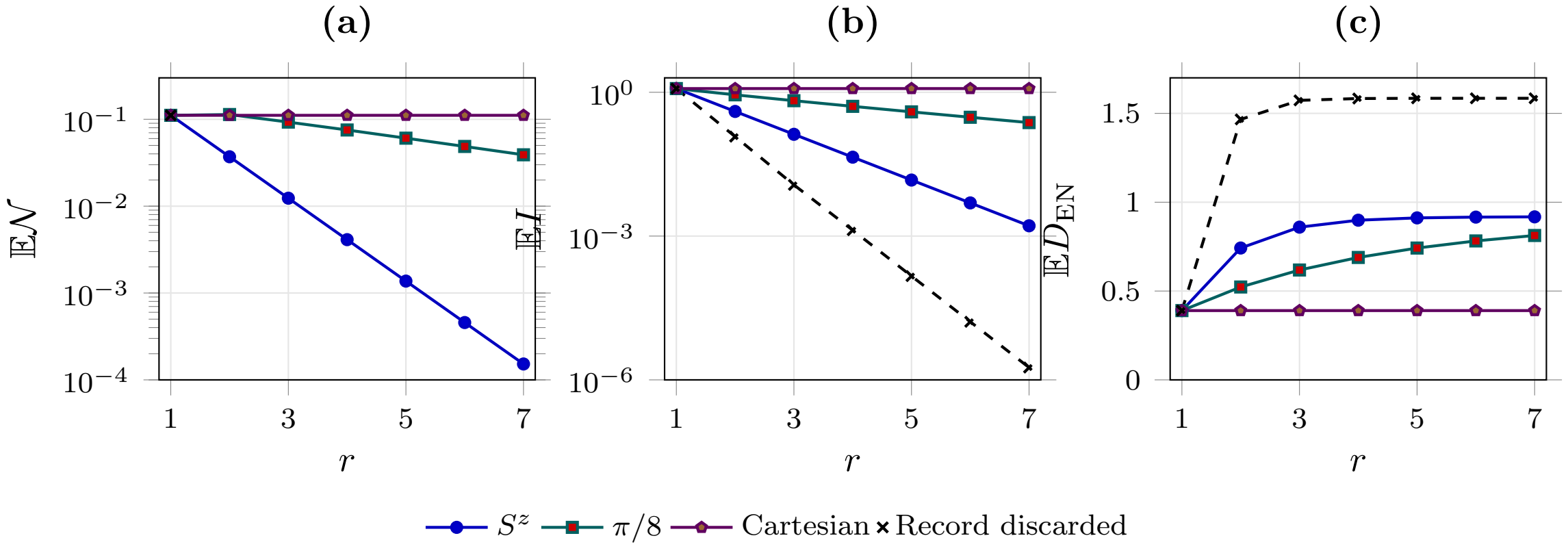
\begin{figure}[htbp]
\centering
\begin{tikzpicture}
\begin{groupplot}[group style={group size=3 by 1,horizontal sep=1.25cm},width=.318\textwidth,height=4.5cm,xlabel={$r$},xtick={1,3,5,7},xmin=.8,xmax=7.2]
\nextgroupplot[title={(a)},ylabel={$\mathbb E\mathcal N$},ymode=log,ymin=0.0001,ymax=.3,unbounded coords=discard,filter discard warning=false,legend to name=pairlegend,legend columns=4]
\addplot+[blue!75!black,mark=*,mark size=1.4pt] table[row sep=\\,x=x,y=N] {
x N I DEN EN\\
1 0.1111111111111 1.19498750024 0.3899750004808 0.289506617195\\
2 0.03703703703704 0.3983291667468 0.7421888895299 0.09650220573166\\
3 0.01234567901235 0.1327763889156 0.859593519213 0.03216740191055\\
4 0.004115226337449 0.0442587963052 0.898728395774 0.01072246730352\\
5 0.001371742112483 0.01475293210173 0.9117733546277 0.003574155767839\\
6 0.0004572473708276 0.004917644033911 0.9161216742455 0.001191385255946\\
7 0.0001524157902759 0.001639214677971 0.9175711141182 0.0003971284186487\\
};
\addlegendentry{$S^z$}
\addplot+[teal!75!black,mark=square*,mark size=1.4pt] table[row sep=\\,x=x,y=N] {
x N I DEN EN\\
1 0.1111111111111 1.19498750024 0.3899750004808 0.289506617195\\
2 0.1134998186017 0.8848258178016 0.522748254783 0.295129347513\\
3 0.09282679187712 0.6676874867903 0.6184157504107 0.2438803659381\\
4 0.07534816636145 0.5088003770317 0.6890413573275 0.1992314153515\\
5 0.06063956172055 0.390380337977 0.7419922038252 0.1611745239489\\
6 0.04867566385111 0.3010440031563 0.7820961131408 0.1299182974098\\
7 0.03902153365328 0.2330757318396 0.8126965018185 0.1045166086159\\
};
\addlegendentry{$\pi/8$}
\addplot+[violet!75!black,mark=pentagon*,mark size=1.4pt] table[row sep=\\,x=x,y=N] {
x N I DEN EN\\
1 0.1111111111111 1.19498750024 0.3899750004808 0.289506617195\\
2 0.1111111111111 1.19498750024 0.3899750004808 0.289506617195\\
3 0.1111111111111 1.19498750024 0.3899750004808 0.289506617195\\
4 0.1111111111111 1.19498750024 0.3899750004808 0.289506617195\\
5 0.1111111111111 1.19498750024 0.3899750004808 0.289506617195\\
6 0.1111111111111 1.19498750024 0.3899750004808 0.289506617195\\
7 0.1111111111111 1.19498750024 0.3899750004808 0.289506617195\\
};
\addlegendentry{Cartesian}
\addplot+[black,only marks,mark=x,mark size=2pt] table[row sep=\\,x=x,y=N] {
x N I DEN\\
1 0.1111111111111 1.19498750024 0.3899750004808\\
2 0 0.1187412029825 1.466221297739\\
3 0 0.01169957450375 1.573262926217\\
4 0 0.001328102146975 1.583634398574\\
5 0 0.0001462987515728 1.58481620197\\
6 0 1.62993752415e-05 1.584946201346\\
7 0 1.809377535356e-06 1.584960691344\\
};
\addlegendentry{Record discarded}
\nextgroupplot[title={(b)},ylabel={$\mathbb E I$},ymode=log,ymin=0.000001,ymax=2]
\addplot+[blue!75!black,mark=*,mark size=1.4pt] table[row sep=\\,x=x,y=I] {
x N I DEN EN\\
1 0.1111111111111 1.19498750024 0.3899750004808 0.289506617195\\
2 0.03703703703704 0.3983291667468 0.7421888895299 0.09650220573166\\
3 0.01234567901235 0.1327763889156 0.859593519213 0.03216740191055\\
4 0.004115226337449 0.0442587963052 0.898728395774 0.01072246730352\\
5 0.001371742112483 0.01475293210173 0.9117733546277 0.003574155767839\\
6 0.0004572473708276 0.004917644033911 0.9161216742455 0.001191385255946\\
7 0.0001524157902759 0.001639214677971 0.9175711141182 0.0003971284186487\\
};
\addplot+[teal!75!black,mark=square*,mark size=1.4pt] table[row sep=\\,x=x,y=I] {
x N I DEN EN\\
1 0.1111111111111 1.19498750024 0.3899750004808 0.289506617195\\
2 0.1134998186017 0.8848258178016 0.522748254783 0.295129347513\\
3 0.09282679187712 0.6676874867903 0.6184157504107 0.2438803659381\\
4 0.07534816636145 0.5088003770317 0.6890413573275 0.1992314153515\\
5 0.06063956172055 0.390380337977 0.7419922038252 0.1611745239489\\
6 0.04867566385111 0.3010440031563 0.7820961131408 0.1299182974098\\
7 0.03902153365328 0.2330757318396 0.8126965018185 0.1045166086159\\
};
\addplot+[violet!75!black,mark=pentagon*,mark size=1.4pt] table[row sep=\\,x=x,y=I] {
x N I DEN EN\\
1 0.1111111111111 1.19498750024 0.3899750004808 0.289506617195\\
2 0.1111111111111 1.19498750024 0.3899750004808 0.289506617195\\
3 0.1111111111111 1.19498750024 0.3899750004808 0.289506617195\\
4 0.1111111111111 1.19498750024 0.3899750004808 0.289506617195\\
5 0.1111111111111 1.19498750024 0.3899750004808 0.289506617195\\
6 0.1111111111111 1.19498750024 0.3899750004808 0.289506617195\\
7 0.1111111111111 1.19498750024 0.3899750004808 0.289506617195\\
};
\addplot+[black,dashed,mark=x,mark size=2pt] table[row sep=\\,x=x,y=I] {
x N I DEN\\
1 0.1111111111111 1.19498750024 0.3899750004808\\
2 0 0.1187412029825 1.466221297739\\
3 0 0.01169957450375 1.573262926217\\
4 0 0.001328102146975 1.583634398574\\
5 0 0.0001462987515728 1.58481620197\\
6 0 1.62993752415e-05 1.584946201346\\
7 0 1.809377535356e-06 1.584960691344\\
};
\nextgroupplot[title={(c)},ylabel={$\mathbb E D_{\rm EN}$},ymin=0,ymax=1.7]
\addplot+[blue!75!black,mark=*,mark size=1.4pt] table[row sep=\\,x=x,y=DEN] {
x N I DEN EN\\
1 0.1111111111111 1.19498750024 0.3899750004808 0.289506617195\\
2 0.03703703703704 0.3983291667468 0.7421888895299 0.09650220573166\\
3 0.01234567901235 0.1327763889156 0.859593519213 0.03216740191055\\
4 0.004115226337449 0.0442587963052 0.898728395774 0.01072246730352\\
5 0.001371742112483 0.01475293210173 0.9117733546277 0.003574155767839\\
6 0.0004572473708276 0.004917644033911 0.9161216742455 0.001191385255946\\
7 0.0001524157902759 0.001639214677971 0.9175711141182 0.0003971284186487\\
};
\addplot+[teal!75!black,mark=square*,mark size=1.4pt] table[row sep=\\,x=x,y=DEN] {
x N I DEN EN\\
1 0.1111111111111 1.19498750024 0.3899750004808 0.289506617195\\
2 0.1134998186017 0.8848258178016 0.522748254783 0.295129347513\\
3 0.09282679187712 0.6676874867903 0.6184157504107 0.2438803659381\\
4 0.07534816636145 0.5088003770317 0.6890413573275 0.1992314153515\\
5 0.06063956172055 0.390380337977 0.7419922038252 0.1611745239489\\
6 0.04867566385111 0.3010440031563 0.7820961131408 0.1299182974098\\
7 0.03902153365328 0.2330757318396 0.8126965018185 0.1045166086159\\
};
\addplot+[violet!75!black,mark=pentagon*,mark size=1.4pt] table[row sep=\\,x=x,y=DEN] {
x N I DEN EN\\
1 0.1111111111111 1.19498750024 0.3899750004808 0.289506617195\\
2 0.1111111111111 1.19498750024 0.3899750004808 0.289506617195\\
3 0.1111111111111 1.19498750024 0.3899750004808 0.289506617195\\
4 0.1111111111111 1.19498750024 0.3899750004808 0.289506617195\\
5 0.1111111111111 1.19498750024 0.3899750004808 0.289506617195\\
6 0.1111111111111 1.19498750024 0.3899750004808 0.289506617195\\
7 0.1111111111111 1.19498750024 0.3899750004808 0.289506617195\\
};
\addplot+[black,dashed,mark=x,mark size=2pt] table[row sep=\\,x=x,y=DEN] {
x N I DEN\\
1 0.1111111111111 1.19498750024 0.3899750004808\\
2 0 0.1187412029825 1.466221297739\\
3 0 0.01169957450375 1.573262926217\\
4 0 0.001328102146975 1.583634398574\\
5 0 0.0001462987515728 1.58481620197\\
6 0 1.62993752415e-05 1.584946201346\\
7 0 1.809377535356e-06 1.584960691344\\
};
\end{groupplot}
\end{tikzpicture}
\par\smallskip\pgfplotslegendfromname{pairlegend}
\caption{Correlations between two physical spins. Colored curves show Born-weighted averages. The discarded-record negativity is zero for $r\ge2$ and is omitted on the logarithmic axis. Cartesian assistance preserves the nearest-neighbor values. In the $S^z$ basis, negativity and mutual information fall by $3^{-(r-1)}$. The third panel shows the different behavior of $D_{\rm EN}$.}
\label{fig:pairs}
\end{figure}

\begin{proposition}[Record-free physical pairs]\label{prop:barepair}
For separation $r\ge1$, set $q=(-1/3)^{r-1}$. In a Cartesian spin basis, write $F$ for the swap on $\mathbb C^3\otimes\mathbb C^3$ and $\ket\Omega=\sum_{a=x,y,z}\ket{aa}$. Then
\begin{equation}
\rho_{i,i+r}=\frac{\one+q(\ket\Omega\!\bra\Omega-F)}{9}.
\label{eq:barepair}
\end{equation}
The nearest-neighbor pair has negativity $1/9$ and logarithmic negativity $\log_2(11/9)$. The pair is separable for every $r\ge2$. Its mutual information has the asymptotic expansion
\begin{equation}
 I(i{:}i+r)=\frac{2}{3\ln 2}\,9^{-(r-1)}
              +O\bigl(27^{-(r-1)}\bigr).
\label{eq:mi_decay}
\end{equation}
\end{proposition}
\begin{proof}
Choose the Cartesian phases so that $A_a=\sigma^a/\sqrt3$. The trace identity for four Pauli matrices gives $\rho_{i,i+1}=(\one+\ket\Omega\!\bra\Omega-F)/9$. Equation~\eqref{eq:depol} writes the general pair as $q\rho_{i,i+1}+(1-q)\one/9$, proving Eq.~\eqref{eq:barepair}. The eigenspaces are the line spanned by $\Omega$, the antisymmetric subspace, and the symmetric subspace orthogonal to $\Omega$. The respective eigenvalues are $(1+2q)/9$, $(1+q)/9$, and $(1-q)/9$, with multiplicities $1,3,5$. Partial transposition swaps $F$ with $\ket\Omega\!\bra\Omega$ and hence sends $q$ to $-q$. This yields the stated nearest-neighbor negativities.

For $r\ge2$, the channel between the target sites contains a factor $\mathcal E$. Its Choi state $J_1$ is separable by Proposition~\ref{prop:edges}, so it is entanglement breaking~\cite{Horodecki2003}. Applying this channel to the joint state of the first emitted spin and the virtual qubit makes that state separable. Emitting the second spin and tracing out the virtual output are local operations on the virtual side and preserve separability. This strengthens the vanishing-negativity conclusion for this pair geometry~\cite{SantosKorepinBose2011,SantosKorepin2012}.

The marginals are $\one/3$, so the mutual information is the relative entropy to $\one/9$. Expanding $(1+x)\ln(1+x)=x+x^2/2+O(x^3)$ at $x=0$ for the three eigenvalues gives a vanishing linear term and a quadratic term $(4+3+5)q^2/(18\ln2)$.
\end{proof}
In particular, $I_{\rm nn}=2\log_2 3-S(J_2)\simeq1.195$ bits and ${\DEN}_{\rm nn}=S(J_2)-\log_2 3\simeq0.390$ bits. The nearest-neighbor state has negative partial transpose and a positive value of $\DEN$.

\section{LOCC-assisted entanglement via the HQMM filter}\label{sec:retained-record}
\subsection{Edge--edge: exact laws}

\begin{theorem}[Filtered edge entanglement]\label{thm:edges}
In the edge--edge scenario with record kept, for every basis $\bb$:
\begin{align}
  &\E[\mathcal N]=\tfrac12\,e^{-\kappa_\bb n},\qquad \E[C]=e^{-\kappa_\bb n},\label{eq:N}\\
  &e^{-2\kappa_\bb n}\le \E[S(\rho_n)]\le e^{-\kappa_\bb n},\label{eq:Sbounds}\\
  &\E[\DEN]=-\E[S(\rho_n)],\qquad \E[I(L{:}R)]=2\,\E[S(\rho_n)],\label{eq:DEN}\\
  &\chi_n+\E[S(\rho_n)]=1,\label{eq:compl}
\end{align}
Here $D$ is the classical record register and $R$ is Bob's edge qubit. Their joint state and mutual information are
\[
  \omega_{DR}=\sum_{d:\,p_d>0}p_d\ket d\!\bra d\otimes\rho_n(d),
  \qquad
  \chi_n=I(D{:}R)_\omega=S(\E\rho_n)-\E[S(\rho_n)],
\]
where $p_d$ is the Born probability of record $d$. Moreover, $\E[E_{\mathcal N}]\le\log_2(1+e^{-\kappa_\bb n})$.
\end{theorem}

\begin{proof}
The conditional $LR$ state is pure with reduced state $\rho_n$ and $\rho_0=\one/2$, so $C=\mathcal M(\rho_n)$ and $\mathcal N=C/2$. Lemma~\ref{lem:law} with $\mathcal M(\one/2)=1$ gives Eq.~\eqref{eq:N}. Write $t=\sqrt{1-C^2}$ and $E(C)=h((1+t)/2)$, with $h$ the binary entropy. Its expansion $1-E(C)=\sum_{k\ge1}t^{2k}/[2k(2k-1)\ln2]$ has positive coefficients summing to one, so $E(C)\ge 1-t^2=C^2$. For $0<C<1$, $E''(C)=(\operatorname{artanh}t-t)/(t^3\ln2)\ge0$, and the chord between $E(0)=0$ and $E(1)=1$ gives $E(C)\le C$. Jensen's inequality $\E[C^2]\ge(\E C)^2$ then gives Eq.~\eqref{eq:Sbounds}. Equation~\eqref{eq:DEN} is the pure-state form of Eq.~\eqref{eq:den}. The classical blocks of $\omega_{DR}$ give $S(\omega_{DR})=H(p)+\E[S(\rho_n)]$, while $S(\omega_D)=H(p)$, where $H(p)$ is the Shannon entropy of the record. Thus $I(D{:}R)_\omega=S(\E\rho_n)-\E[S(\rho_n)]$. Since $\mathcal E$ is unital, $\E\rho_n=\one/2$ and $S(\E\rho_n)=1$, proving Eq.~\eqref{eq:compl}. The bound on $E_{\mathcal N}=\log_2(1+C)$ follows from concavity.
\end{proof}

\begin{corollary}[Localizable entanglement]
For the cut $L|CR$, $\sup_\bb\E[S]\le1$ ebit. The bound is attained at every distance $n$ by any real Cartesian frame, for which $\kappa_\bb=0$. The edge entanglement length is therefore infinite, while $\xi=1/\ln3$. In the $S^z$ basis, $\E[S]=3^{-n}$ exactly.
\end{corollary}

\begin{figure}[htbp]
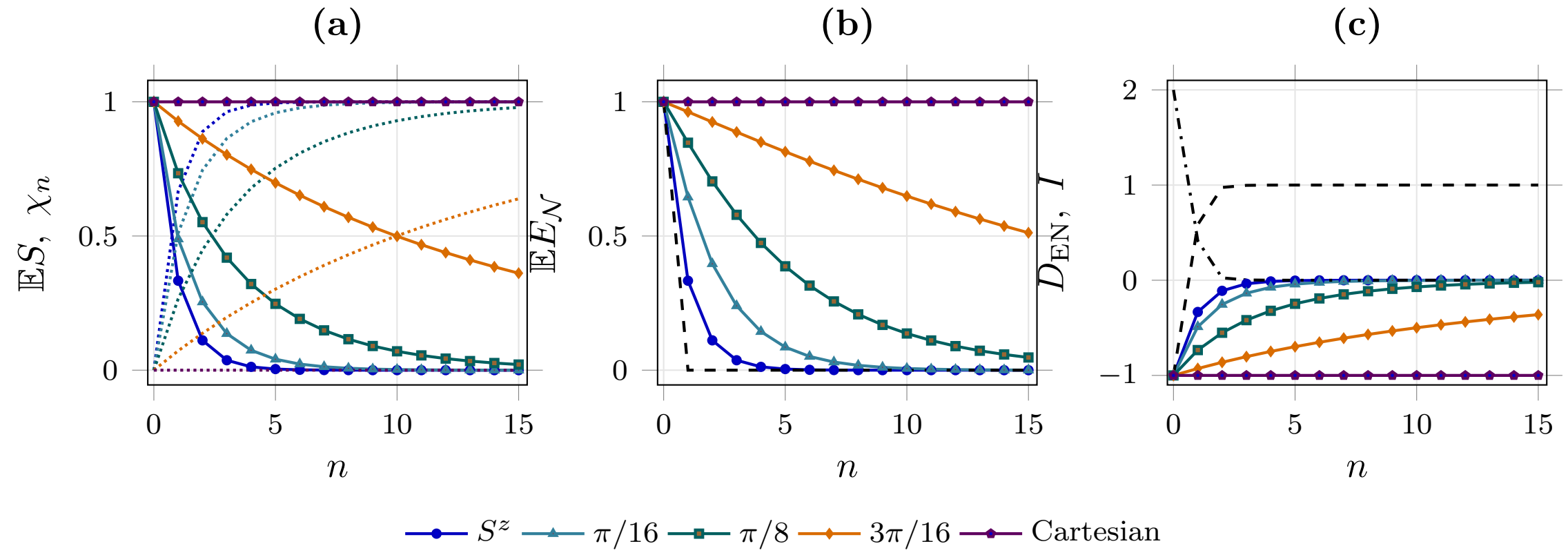

\centering

\par\smallskip\pgfplotslegendfromname{edgelegend}
\caption{Edge-pair observables from the finite record sum, starting with $\rho_0=\one/2$. (a) Entropy (solid) and Holevo information (dotted). The two curves for each basis sum to one. (b) Mean logarithmic negativity with the record (colored) and logarithmic negativity of the discarded-record state (dashed). (c) Conditional mean $D_{\rm EN}$ (colored) and record-free $D_{\rm EN}$ (dashed) and mutual information (dash-dotted). The discarded-record state is separable for $n\ge1$.}
\label{fig:edges}
\end{figure}

This recovers the diverging entanglement length of Verstraete, Mart\'in-Delgado, and Cirac~\cite{VerstraeteMartinDelgadoCirac2004}, now as the $\kappa_\bb=0$ point of a continuous family. By Nielsen's theorem~\cite{Nielsen1999,vanLuijk2025CMP}, the maximally entangled pure edge state obtained in the Cartesian basis can be converted by LOCC into any two-qubit pure state. Its Schmidt rank, $2$, is bounded by the bond dimension~\cite{vanLuijk2024CMP}. Equation~\eqref{eq:compl} states that the quantum mutual information of the classical record and Bob's edge plus the mean conditional entanglement entropy equals one bit. If a measurement on $R$ produces a classical outcome $Y$, data processing gives $I(D{:}Y)\le\chi_n$~\cite{Holevo1973}. For the real one-parameter family described after Eq.~\eqref{eq:filter}, each Kraus matrix is diagonal or antidiagonal in the virtual $\sigma^z$ basis. Starting from $\one/2$, every conditional $\rho_n(d)$ is therefore diagonal in that same basis. Measuring $R$ in it preserves all eigenvalue probabilities of the classical--quantum state, so $I(D{:}Y)=\chi_n$. Figure~\ref{fig:edges} compares the retained-record averages with the record-free curves of Proposition~\ref{prop:edges}.

For each record $d$ of positive probability, the conditional edge pair is pure and
\[
  \DEN(\rho_{LR}^{(d)})=-S(\rho_n(d))\le 0.
\]
Since the initial virtual state has full rank, $\rho_n(d)$ has rank $\operatorname{rank}K_d$. The inequality is strict precisely when $K_d$ has rank two. For example, at $n=1$ in the $S^z$ basis, each outcome has probability $1/3$. Outcomes $+$ and $-$ give product states with $\DEN=0$, and outcome $0$ gives a Bell state with $\DEN=-1$. The mean is strictly negative for every finite $n$, because Eqs.~\eqref{eq:Sbounds} and~\eqref{eq:DEN} give $\E[\DEN]\le -s_\bb^{2n}<0$. For $n\ge1$, discarding the record gives $\DEN(J_n)\ge0$ and $\mathcal N(J_n)=0$ by Proposition~\ref{prop:edges}.

\subsection{Spin--spin: distance-independent entanglement}

\begin{theorem}[Cartesian assistance]\label{thm:cart}
Let $\bb$ be any real Cartesian frame. Every record of the $r-1$ middle spins has probability $3^{-(r-1)}$, and
\begin{equation}
  \rho^{(\mathcal D)}_{i,i+r}=\bigl(D(g_{\mathcal D})\otimes\one\bigr)\,\rho_{i,i+1}\,\bigl(D(g_{\mathcal D})\otimes\one\bigr)^\dagger,
  \label{eq:LU}
\end{equation}
where $g_{\mathcal D}$ is the identity or a $\pi$ rotation about a measured Cartesian axis, and $K_{\mathcal D}\propto u_{g_{\mathcal D}}$. In Eq.~\eqref{eq:LU}, $D(g_{\mathcal D})$ on the physical output is understood in a Cartesian basis adapted to the frame. Hence every measure $f$ that is invariant under local unitaries ($\mathcal N$, $E_{\mathcal N}$, $I$, $\DEN$) satisfies $\E f(\rho^{(\mathcal D)}_{i,i+r})=f_{\rm nn}$ for all $r$.
\end{theorem}

\begin{proof}
For a Cartesian frame, each $K_a$ is a phased Pauli along the corresponding axis divided by $\sqrt3$. Thus $K_{\mathcal D}=c\,3^{-(r-1)/2}u_g$ with $|c|=1$ and $u_g$ a product of such Pauli matrices. The probability follows from $p(\mathcal D)=\tr(K_{\mathcal D}K_{\mathcal D}^\dagger)/2$. Inserting $u_g$ into Eq.~\eqref{eq:pair} conjugates $A_s$ by $u_g$ at the first target. Equation~\eqref{eq:cov} acts on its output coefficient column by $D(g)^T$, which equals $D(g)$ in this adapted frame since the rotation has order at most two and the Cartesian representation is real orthogonal. The right target is unchanged.
\end{proof}

\begin{proposition}[$S^z$ assistance]\label{prop:sz}
In the $S^z$ basis, $\E f(\rho^{(\mathcal D)}_{i,i+r})=f_{\rm nn}\,3^{-(r-1)}$ for every measure $f$ that is invariant under local unitaries and vanishes on product states. In particular, $\E\mathcal N=\tfrac19\,3^{-(r-1)}$.
\end{proposition}

\begin{proof}
If any middle outcome is $\pm$ and the record has positive probability, then $K_{\mathcal D}$ has rank one, $K_{\mathcal D}=\ket a\!\bra b$. The trace in Eq.~\eqref{eq:pair} then factorizes into an $(s,s')$ part times a $(t,t')$ part, giving a product state. The all-zero record has probability $3^{-(r-1)}$ because $A_0^\dagger A_0=\one/3$. For that record, $K_{\mathcal D}\propto(\sigma^z)^{r-1}$, a $D_2$ element, and Theorem~\ref{thm:cart} applies.
\end{proof}

The product branches in the $S^z$ protocol have zero mutual information and zero negativity. Their $\DEN$ equals the mean of their two marginal entropies, so Proposition~\ref{prop:sz} gives no analogous geometric scaling for $\DEN$. The Cartesian-assisted pair retains ${\DEN}_{\rm nn}>0$ at every separation, despite having negativity $1/9$.

\subsection{A non-Cartesian basis with larger negativity}
\begin{proposition}\label{prop:enhancement}
For the real basis family at $\theta=\pi/6$ and $r=2$, the mean negativity is
\begin{equation}
  \E\mathcal N=\frac{\sqrt7+\sqrt3-1}{27}
     >\frac19.
\end{equation}
\end{proposition}
\begin{proof}
Each one-site outcome has probability $1/3$. The $0$ outcome is a virtual Pauli and contributes $1/27$ to the mean. The two other Kraus matrices at this angle are related by virtual unitaries on their left and right: if $K_1=-a\sigma^+-b\sigma^-$ and $K_2=b\sigma^+-a\sigma^-$, then $K_2=-\sigma^z\sigma^x K_1\sigma^x$. For virtual unitaries $U,V$ and $K'=UKV$, the two-target amplitude factors as
\[
  A_tK'A_s=U\,(U^\dagger A_t U)\,K\,(VA_sV^\dagger)\,V.
\]
In Eq.~\eqref{eq:pair}, the outer $V$ cancels at the maximally mixed input $\one/2$, and the outer $U$ cancels by cyclicity of the trace. Up to phases, $U$ and $V$ are lifts of rotations $g_U$ and $g_V$. Equation~\eqref{eq:cov} converts $VA_sV^\dagger$ into a rotation on the first target and $U^\dagger A_t U$ into a rotation on the second. Explicitly, the matrices acting on the physical coefficient indices $s,t$ are $D(g_V)^T$ and $D(g_U^{-1})^T$, respectively, and both are unitary. Consequently the unnormalized pair operators are locally unitarily conjugate and have the same trace. The two outcomes thus have equal negativity.

For outcome $1$, the unnormalized partial transpose in the $S^z$ basis is block diagonal after reordering. Multiplying it by $27$ gives scalar blocks $1,3$ and the blocks
\[
.
\]
The blocks follow by substituting $K_1=-\sqrt3 A_+/2+A_-/2$ into Eq.~\eqref{eq:pair} and transposing the second physical index. The negative eigenvalues of the scaled matrix are $(1-\sqrt3)/2$ and $(1-\sqrt7)/2$. The absolute sum of the negative eigenvalues of the unnormalized partial transpose already includes the Born weight. Adding the two equal contributions and the $0$ outcome proves the equality. The strict comparison follows from $\sqrt7+\sqrt3>4$.
\end{proof}

\begin{figure}[htbp]
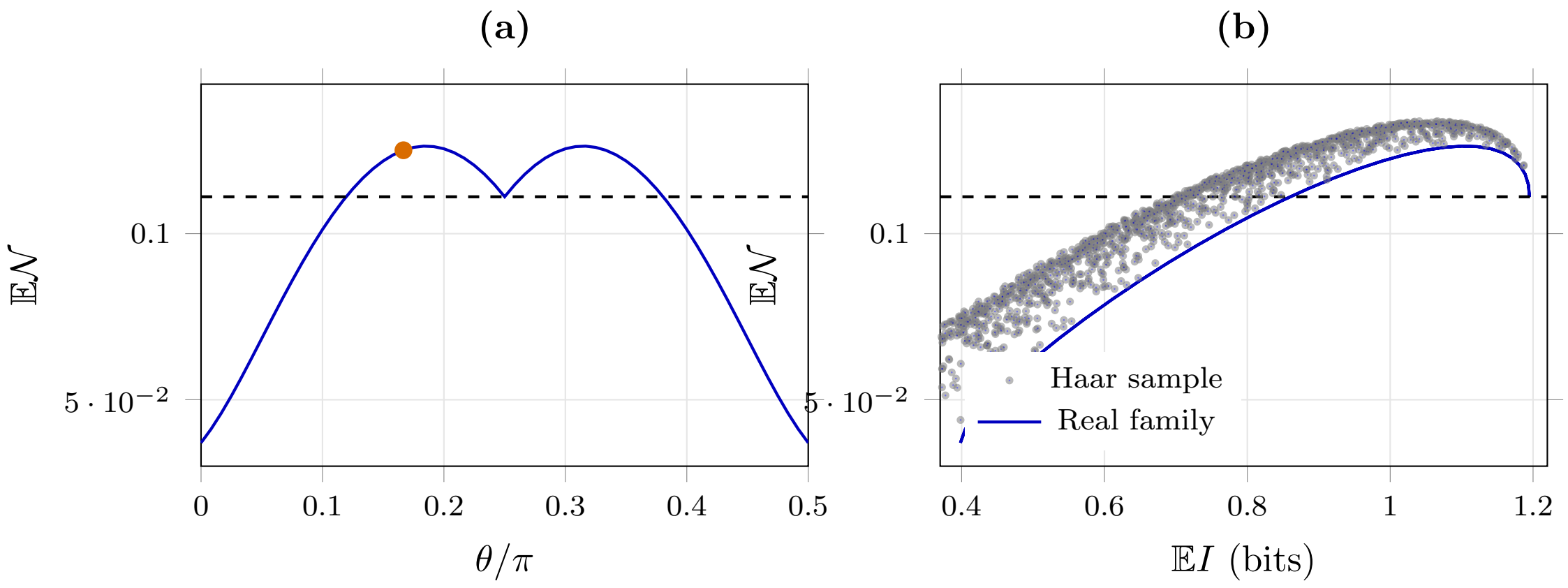

\centering
%

\caption{One-site basis scan for the physical pair at $r=2$. (a) Average negativity for the real family. The highlighted point is the exact example of Proposition~\ref{prop:enhancement}. (b) Mean negativity versus mean mutual information for the angular grid and the fixed Haar sample. The dashed line marks $\mathcal N_{\rm nn}=1/9$.}
\label{fig:scan}
\end{figure}

The scan in Fig.~\ref{fig:scan} uses a 61-point angular grid and a fixed sample of 1500 Haar-distributed bases in $U(3)$. The largest negativity values in these sets are approximately $0.12634$ and $0.13386$, respectively. Their respective largest mutual information values are $1.19499$ and $1.18702$ bits. These are finite-sample comparisons. Proposition~\ref{prop:enhancement} establishes non-optimality of Cartesian measurements for this spin-pair negativity problem independently of the numerical scan.

\subsection{Figure data}
For Fig.~\ref{fig:edges}, the real-family Kraus matrices are diagonal or antidiagonal. Set $a=2\cos^2\theta/3$ and $b=2\sin^2\theta/3$. The diagonal entries of a branch at depth $n$ are $(a^u b^v,a^v b^u)/(2\,3^w)$ with $u+v+w=n$. Multiplicities are propagated by the updates $(u,v)\mapsto(u,v)$, $(v+1,u)$ and $(v,u+1)$, with the new depth determining $w$. This groups all records into a finite weighted sum. Figure~\ref{fig:pairs} uses direct enumeration of the $3^{r-1}$ records and Born-weighted spectral calculations on the corresponding $9\times9$ states. The figures use floating-point evaluation of these sums. The numerical data are included in the standalone LaTeX source.

\section{Discussion}\label{sec:discussion}
Figure~\ref{fig:setup} distinguishes two uses of the measurement record. For edge qubits measured in a Cartesian basis, it specifies a Pauli correction that standardizes the distributed Bell pair. For physical spin targets, it labels the conditional ensemble used to evaluate mean negativity. The exact distance-two example gives a benchmark for comparing measurement bases in an entanglement-distribution experiment. At the edges, the determinant scale $s_\bb$ fixes the mean concurrence and negativity. A Cartesian measurement preserves one ebit along every record, recovering the perfect-wire protocol~\cite{VerstraeteMartinDelgadoCirac2004,GrossEisert2007,BrennenMiyake2008}. The Holevo identity equates the reduction from one ebit in the mean conditional entanglement to the quantum mutual information between the record and Bob's edge. For the physical-spin pair, the two emission steps produce a mixed state. In this geometry, a nonunitary virtual update can increase mean negativity above the value obtained by a Pauli correction.

The entropic and partial-transpose tests capture different features. For the flagged edge ensemble, the average conditional pure-state entropy is achievable as an asymptotic entanglement rate: for independent preparations with fixed $n$, the relative frequency of each positive-probability record $d$ converges to $p_d$. The parties can use the broadcast records to sort the pairs and concentrate each pure-state subensemble at its entropy rate $S(\rho_n(d))$, giving a total rate $\sum_d p_d S(\rho_n(d))$ ebits per preparation. With the flag discarded, the edge state is separable. For the physical-spin pair, the Cartesian protocol distributes a negative-partial-transpose state whose positive $\DEN$ gives an inconclusive hashing test. This illustrates how the choice of target systems affects the interpretation of entanglement measures.

The transfer-channel and filter derivations connect the finitely correlated description of AKLT~\cite{Fannes1992,Cirac2021} with its HQMM formulations~\cite{SouissiAndolsiEPJP2026,SouissiAndolsi2026causal,SouissiJSP2025}. The distinction between the projective symmetry of the virtual edge and the specific Kraus matrices at the AKLT point is relevant for MBQC throughout the Haldane phase~\cite{Else2012,Raussendorf2017}. On experimental platforms where AKLT states are prepared with mid-circuit readout and feedforward~\cite{Smith2023}, the exact concurrence and negativity laws provide basis-dependent benchmarks for edge-state reconstruction.

\FloatBarrier
\begin{acknowledgments}
The author acknowledges support from Tunis El Manar University.
\end{acknowledgments}

\end{document}